\documentclass[amsmath,amssymb,prl,reprint]{revtex4-2}
\usepackage{graphicx}
\usepackage{dcolumn}
\usepackage{bm}
\usepackage{natbib}
\usepackage{color}
\usepackage{amsmath}
\usepackage[]{hyperref}
\hypersetup{
    colorlinks=true,
    allcolors = {blue},
    }
\usepackage[normalem]{ulem}
\newcommand{\add}[1]{\textcolor{black}{#1}}

\usepackage{scalerel,stackengine}
\stackMath
\newcommand\reallywidehat[1]{%
\savestack{\tmpbox}{\stretchto{%
  \scaleto{%
    \scalerel*[\widthof{\ensuremath{#1}}]{\kern-.6pt\bigwedge\kern-.6pt}%
    {\rule[-\textheight/2]{1ex}{\textheight}}
  }{\textheight}%
}{0.5ex}}%
\stackon[1pt]{#1}{\tmpbox}%
}
\begin{document}

\title{Quantum Hall Bogoliubov Interferometer}
\author{Vadim~Khrapai}
\affiliation{Osipyan Institute of Solid State Physics, Russian Academy of
Sciences, 142432 Chernogolovka, Russian Federation}
\affiliation{National Research University Higher School of Economics, 20 Myasnitskaya Street, 101000 Moscow, Russian Federation}

\begin{abstract}
A quantum Hall interferometer containing a grounded superconducting terminal is proposed. This geometry allows to control the Andreev and normal scattering amplitudes of sub-gap Bogoliubov quasiparticles with the Aharonov-Bohm phase, as well as with the constrictions defining the interferometer loop. The conductance matrix of \add{such} a three-terminal \add{NSN interference device} exhibits a much richer behavior as compared to its two-terminal Fabry-P\'erot counterpart, \add{which is illustrated by non-trivial behavior of non-local charge and heat responses}. \add{A single edge version of the interferometer enables full on-demand control of the electron-hole superposition, including resonant enhancement of arbitrary small Andreev reflection probability up to 1, and can be used as a building block in future more complex interference setups.}
\end{abstract}

\maketitle

Chiral one-dimensional transport of quasiparticles along the boundary of a gapped two-dimensional electron system (2DES) is a fundamental aspect of the quantum Hall effect~\cite{Halperin_1982,Buettiker_1988}. A combination of phase-coherent ballistic propagation over large distances together with a controllable backscattering by gate-defined constrictions result in a plethora of quasiparticle interference phenomena in quantum Hall edge channels~\cite{Carrega_2021}. Matching these unique capabilities with a superconducting proximity effect may greatly advance the research in semiconductor-superconductor hybrids.  

A semi-classical transport of Bogoliubov quasiparticles by skipping orbits along 2DES-superconductor interface has been realized in Refs.~\cite{Eroms_2005,Eroms_2007,Batov_2007}. In this low magnetic field range the Andreev reflection process, which is constrainted by momentum conservation, is allowed for scattering between different edge modes or in the presence of disorder~\cite{Chtchelkatchev_2001,Giazotto_2005,Chtchelkatchev_2007}. More recently, the experiments extended towards the physics of chiral Andreev edge states in the quantum Hall regime~\cite{Rickhaus_2012,Wan_2015,Kozuka_2018,Matsuo_2018}. Observations of a small but finite cross-Andreev signals in non-local conductance measurements evidence the Andreev reflection within a single chiral edge mode in graphene~\cite{Zhao_2020,Guel_2022}, that may result from pecularities of a valley spectrum at the edge~\cite{Akhmerov_2007} or, most likely, from strong disorder scattering inherent to real superconductors~\cite{Kurilovich_2022_Disorder}. Quantum interference effects play an important role in quasiparticle transport along the 2DES-superconductor interface~\cite{Chtchelkatchev_2007,Khaymovich_2010,Park_2014,Zhao_2020}. Two-particle interference effects were discussed in devices combining quantum Hall edge channels and superconductivity~\cite{Beenakker_2014,Ferraro_2015}. 

\begin{figure}[h]
\begin{center}
\vspace{0mm}
  \includegraphics[width=1\linewidth]{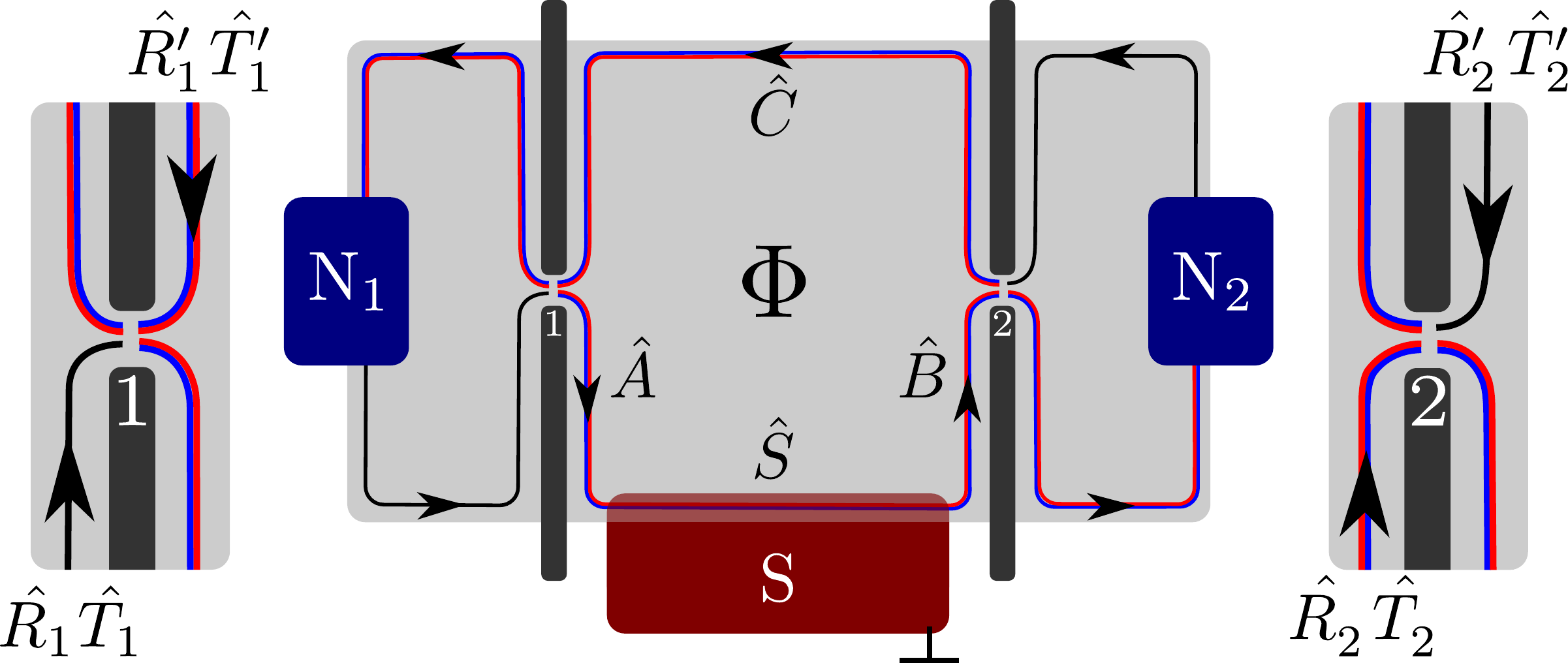}
\end{center}
  \caption{Sketch of the Bogoliubov interferometer. A light-grey rectangle depicts a mesa of the 2DES, with two normal terminals one the sides and a superconducting terminal in the middle, marked, respectively, by $\rm N_1,\,N_2$ and $S$. The edge channel forms a loop of the interferometer with the help of two constrictions, its chirality is shown by arrows. Magnified views of the constrictions 1 and 2 are given on the left and right hand sides. The transmission matrices used in the calculation are shown nearby the corresponding scattering regions.} 
	\label{Fig1}
\end{figure}

In this work, I propose a quantum Hall Bogoliubov interferometer, that is a modification of a Fabry-P\'erot type interferometer~\cite{Chamon_1997,Halperin_2011} with a superconducting terminal inside. This geometry enables a fine tuning of the \add{local and non-local normal and} Andreev scattering amplitudes by means of the Aharonov-Bohm (AB) phase and constrictions defining the interferometer loop. \add{The proposed interferometer represents a versatile three-terminal NSN device that enables full control over the non-local charge and heat quasiparticle transport, including} a resonant enhancement of an arbitrarily small Andreev reflection probability up to 1.

A sketch of the Bogoliubov interferometer is depicted in Fig.~\ref{Fig1}a. The light-grey rectangle represents a mesa of the 2DES, which is divided in three regions separated by  two gate-defined constrictions (schematized by pairs of dark-grey vertical rectangles). The inner region is a Fabry-P\'erot-like interferometer for the chiral edge channels, an essential novel part of which is the grounded superconducting terminal $\mathrm S$ (dark-red). The outer regions on either side of the interferometer contain normal terminals $\mathrm N_1$ and $\mathrm N_2$, which are assumed to be ideally coupled to the chiral edge channels. The edge channels propagating downstream the normal terminals contain normal quasiparticles and are shown in black. Passing the S-terminal, quasiparticles experience both Andreev and normal scattering and become coherent superpositions of the electron-like and hole-like excitations (Bogoliubov quasiparticles), as illustrated by the blue-red color. The whole structure is placed in a quantizing perpendicular magnetic field corresponding to the filling factor $\nu=2$ \add{(the lowest spin-degenerate Landau level filled)}. The phase $\Phi$ is tuned by  the AB flux through the interferometer.

Following Ref.~\cite{Kurilovich_2022_Disorder}, below I assume spin-degeneracy of the chiral edge states and neglect possible edge reconstruction. \add{Hence, all scattering matrices are spin-degenerate and the spin index is suppressed for brevity.}  The calculations are performed at zero energy, that is for quasiparticles at the chemical potential of the S terminal, well inside the bulk and superconductor  gaps. The wave-function of a Bogoliubov quasiparticle is a two-component vector $(a_e\,\,a_h)^\mathrm{T}$, where $a_e$, $a_h$ are the amplitudes of the electron-like and hole-like components, respectively. The propagation around the interferometer is described by $2\times2$ matrix that takes the sum of the amplitudes of all possible trajectories. For example, the transmission matrix corresponding to the entrance via constriction 1 and a single full-turn around the interferometer, is expressed as $\hat{M_0}\hat{T_1}$, where $\hat{M_0} = \hat{R'_1}\hat{C}\hat{R_2}\hat{B}\hat{S}\hat{A}$. Using the scattering amplitudes I express $\hat{A} = \mathrm{diag}(e^{i\phi_A},e^{-i\phi_A})$ (and similar for $\hat{B},\,\hat{C}$) and $\hat{R_i} = \mathrm{diag}(r_i,r_i^*);\,\hat{T_i} = \mathrm{diag}(t_i,t_i^*)$  (and similar for $\hat{R'_i},\,\hat{T'_i}$). 
%
%
Matrices $\hat{A}$, $\hat{B}$ and $\hat{C}$ describe a free propagation along the edge that results only in a phase accumulation. Matrices $\hat{R_i}$ and  $\hat{T_i}$ describe a reflection and transmission of a quasiparticle incident on the $i$-th constriction from the lower edge.  $\hat{R'_i}$ and $\hat{T'_i}$ describe the same processes for a quasiparticle incident from the upper edge (cf. Fig.~\ref{Fig1}). \add{$\hat{R_i}$, $\hat{T_i}$, $\hat{R'_i}$ and $\hat{T'_i}$ are $2\times2$ blocks of the total $4\times4$ scattering matrix of the $i$-th constriction and are constrained by its unitarity~\cite{Blanter_2000}. Therefore, $|r_i|=|r'_i|=\sqrt{R_i}$, $|t_i|=|t'_i|=\sqrt{T_i}$, $r_i^*t_i'= -r_i't_i^*$, where $T_i$ and $R_i$ are, respectively, the transmission and reflection probabilities and $T_i+R_i=1$.} Finally, the only non-diagonal matrix $\hat{S}$ describes a propagation along the 2DES-superconductor boundary. Its non-diagonal elements generate rotations in the electron-hole basis owing to the Andreev scattering:
\begin{equation*}
\hat{S} = \begin{pmatrix}
t_{ee} & t_{eh} \\
t_{he}& t_{hh}
\end{pmatrix}; \,t_{hh} = t_{ee}^*; \,t_{eh} = -t_{he}^*.
\end{equation*}

A detailed microscopic analysis of the matrix $\hat{S}$ was recently performed in Ref.~\cite{Kurilovich_2022_Disorder}. In the following I neglect possible quasiparticle loss in the superconductor, so that $|t_{ee}|^2+|t_{eh}|^2=1$. The full scattering amplitudes are contained in the blocks \add{$\reallywidehat{i\rightarrow j}$}:
%
%
%
\begin{equation*}
\reallywidehat{i\rightarrow j} \equiv 
\begin{pmatrix}
s_{ji}^{ee} & s_{ji}^{eh} \\
 & \\
s_{ji}^{he} & s_{ji}^{hh}
\end{pmatrix},\vspace{2mm}
\end{equation*}
where \add{$i,j\in{1,2}$ label the N terminals} and $s_{ji}^{\alpha\beta}$ represents a scattering amplitude of a quasiparticle of the type $\beta$ from terminal $\mathrm N_i$ to a quasiparticle of the type $\alpha$ in the terminal $\mathrm N_j$ ($\alpha,\beta\in{e,h}$)\add{, see Ref.~\cite{Anantram_1996}}. \add{For example:}
\begin{equation*}
		\reallywidehat{1\rightarrow 2} = \hat{T_2}\hat{B}\hat{S}\hat{A}(1-\hat{M_0})^{-1} \hat{T_1}, %
\end{equation*} 
\add{where} ${(1-\hat{M_0})^{-1}=(1+\hat{M_0}+\hat{M_0}^2+\ldots)}$. Without the loss of generality, the phases accumulated by a quasiparticle before entering the interferometer and after leaving it are assumed zero, so that the corresponding evolution is given by identity matrices. 

\add{The transmission coefficients, defined as $T_{ij}^{\alpha\beta} = |s_{ij}^{\alpha\beta}|^2$, are used to calculate the conductance matrix $G_{ij}\equiv \partial I_i/\partial V_j$. Here, $I_i$ is the electric current flowing in the device through terminal $\mathrm{N}_i$ and $V_j$ is the voltage bias on terminal $\mathrm{N}_j$.} \add{$G_{ij} = G^0\left(\delta_{ij}-T_{ij}^{ee}+T_{ij}^{eh}\right)$~\cite{Anantram_1996}, where $\delta_{ij}$ is the Kronecker delta symbol and $G^0=2e^2/h$ (here 2 stands for the spin degeneracy):}
\begin{widetext}
\begin{equation}
\add{\hat{G} = G^0\frac{T_1 T_2}{D} \begin{pmatrix}
1 & -1 \\
 & \\
-1 & 1
\end{pmatrix} + G^0\frac{2|t_{eh}|^2}{D^2} \begin{pmatrix}
T_1^2R_2 & T_1T_2R_1R_2 \\
 & \\
T_1T_2 & T_2^2R_1
\end{pmatrix},\,D\equiv1+R_1R_2-2\sqrt{R_1R_2}|t_{ee}|\cos{\Phi}} \label{G_matrix}
\end{equation}
\end{widetext}
where $\Phi = \phi_A+\phi_B+\phi_C+\arg(r'_1r_2t_{ee})$, is the phase accumulated by an electron during one full turn around the interferometer. The increment of the AB phase is related to change of a magnetic field $\delta B$ by $\delta\Phi = 2\pi\delta B A/\Phi_0$, where $A$ is the area enclosed by the interferometer and $\Phi_0= h/e$ is the flux quantum. 

\begin{figure}[t]
\begin{center}
\vspace{0mm}
  \includegraphics[width=1\linewidth]{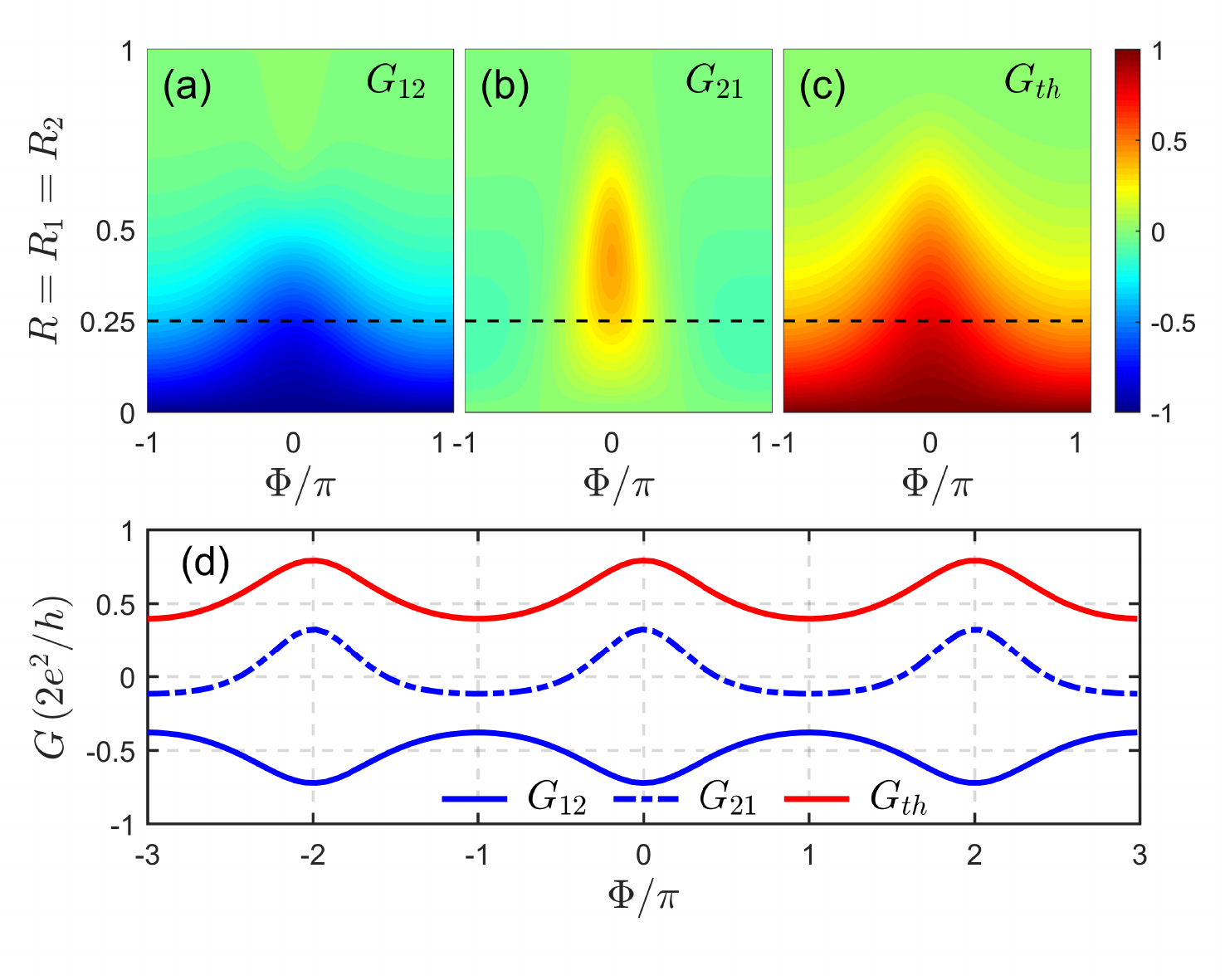}
	\end{center}\vspace{-5mm}
  \caption{\add{Non-local charge and heat responses. (a-c) Color-scale plots of non-local electrical and thermal conductances $G_{12}, G_{21}$ and $G_{th}$, normalized, respectively, by $G^0$ and $G_{th}^0$, as a function of reflection probabilities of the constrictions and the AB phase. The colorbar is common for all panels. (d) Cross--cuts along the dashed lines in panels (a-c), corresponding to $R_1=R_2=0.25$. All data for $|t_{ee}|^2=|t_{eh}|^2=1/2$.}} 
	\label{Fig2} 
	\end{figure}
	
\add{Equation~(\ref{G_matrix}) shows how the Andreev amplitude $t_{eh}$ impacts the conductance matrix of the Bogoliubov interferometer. For $t_{eh} = 0$  the current in the superconductor vanishes and the usual result for a Fabry-P\'erot interferometer is recovered. In this case, the current conservation constraints $\hat{G}$ so that all conductances have the same absolute value, see the first term in Eq.~(\ref{G_matrix}). In the presence of Andreev scattering the S terminal comes into play and the Bogoliubov interferometer becomes a versatile three-terminal NSN device. This can be observed via $\Phi$-controlled non-local charge and heat transport illustrated in Fig.~\ref{Fig2}. Here, apart from $G_{12}$ and $G_{21}$, another quantity of interest is plotted, namely the finite temperature ($\mathrm{T}$) thermal conductance $G_{th}/G_{th}^0\equiv T_{12}^{ee}+T_{12}^{eh}= T_1T_2/D$, where $G_{th}^0 = \mathcal{L}\mathrm{T}G^0$ and $\mathcal{L}$ is the Lorenz number. Figs.~\ref{Fig2}a-\ref{Fig2}c show the color-scale plots of the normalized $G_{12}$, $G_{21}$ and $G_{th}$ as a function of reflection probabilities $R$ of the constrictions (assumed identical) and $\Phi$, for the case of $|t_{eh}|^2=1/2$. All three non-local responses demonstrate pronounced AB oscillations with the amplitude controlled by $R$. Fig.~\ref{Fig2}d details the $\Phi$-dependencies for $R=0.25$ (see the dashed lines in Figs.~\ref{Fig2}a-\ref{Fig2}c). Zeros of $G_{21}(\Phi)$ are accompanied by finite $G_{th}$ and manifest the AB phase tuned non-local charge-heat separation~\cite{Denisov_2021,Denisov_2022}. In addition, the setup of Fig.~\ref{Fig1} can be used to generate entanglement via Cooper pair splitting in the bias regime $V_1=V_2>0$, achieving the splitting efficiency of 50\% for $R_1=R_2\rightarrow1$. Finally, an intriguing feature of the interferometer is that the magnitude of the AB oscillations is determined by the normal scattering amplitude  $|t_{ee}|$, see Eq.~(\ref{G_matrix}), and vanishes in the limit of ideal Andreev reflection.}
\begin{figure}[t]
\begin{center}
\vspace{2mm}
\hspace{5mm}\includegraphics[width=0.8\linewidth]{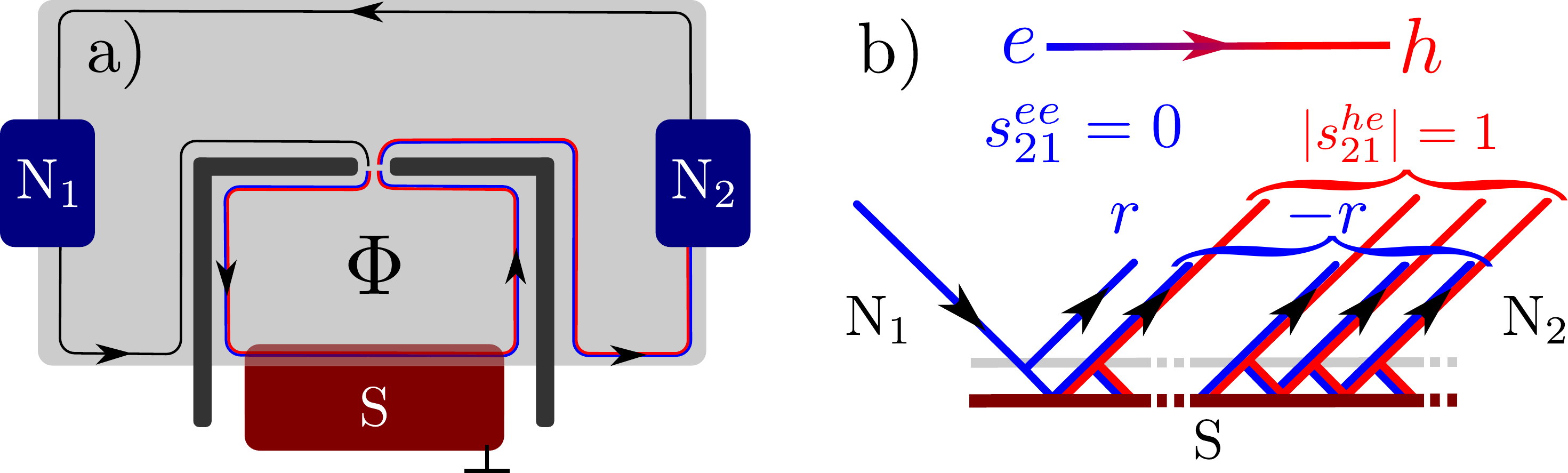}
  \includegraphics[width=0.9\linewidth]{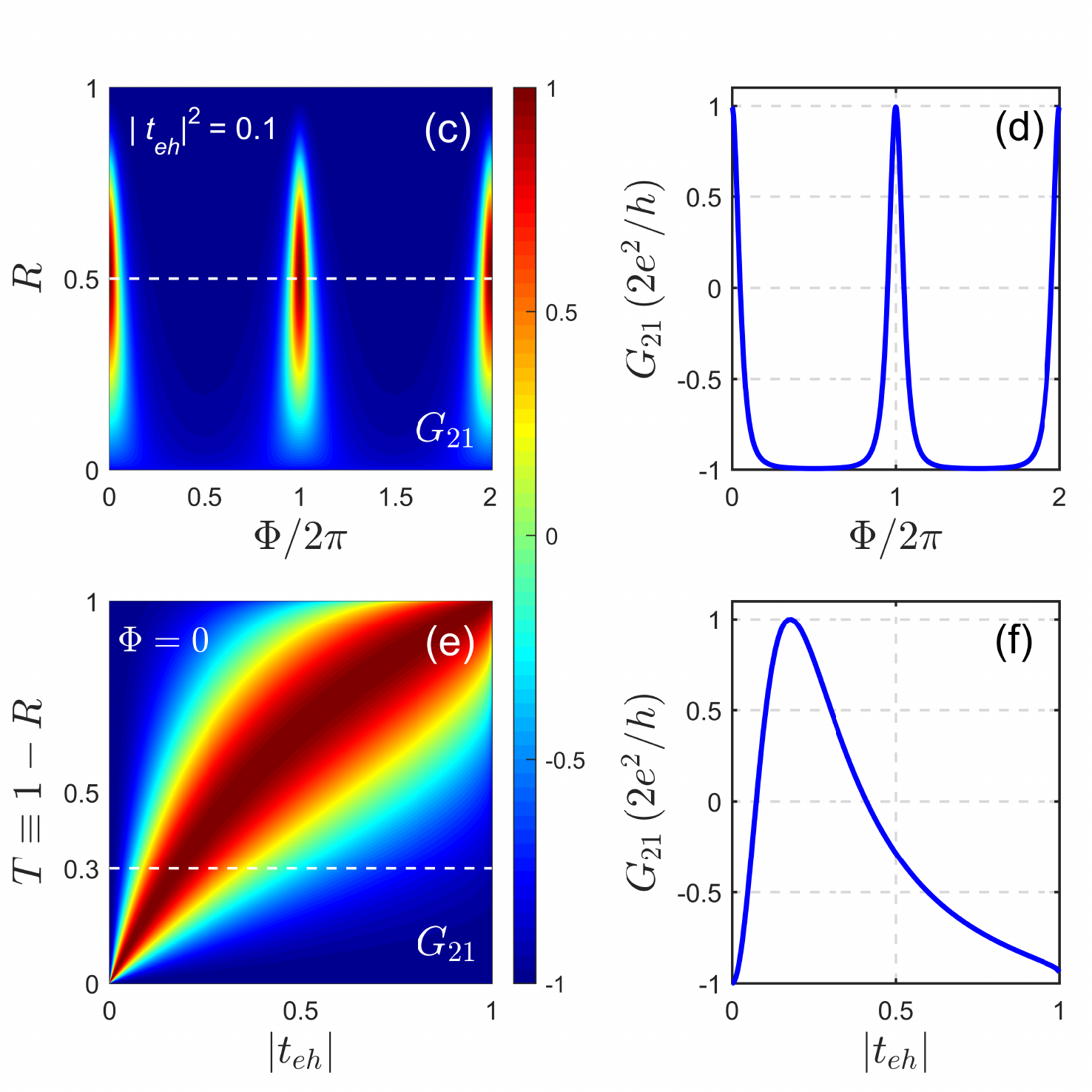}
\end{center}\vspace{-5mm}
  \caption{Bogoliubov interferometer coupled to a single edge. (a) A sketch of the interferometer. (b) Enhancement of a cross-Andreev reflection probability to 1 for $T=T^{\mathrm{opt}}$ (see text). Light-grey (dark-red) lines indicate scattering at the constriction (at the 2DES-superconductor interface). Blue (red) lines are the interfering paths for the electron (hole) wave function components. (c) Color-scale plot of $G_{21}$ on the $\Phi-R$ plane for  $|t_{eh}|^2=0.1$. (d) Cross-cut along the dashed line in panel (c) at $R=0.5$ that demonstrates a resonant enhancement of the non-local conductance. (e) Color-scale plot of $G_{21}$ on resonance ($\Phi=0$) as a function of transmission probability of the contriction and the Andreev amplitude. (f) Cross-cut along the dashed line in panel (e) at $T=0.3$. The colorbar in units of $2e^2/h$ is common for (c) and (e).} 
	\label{Fig3} 
\end{figure}

Bogoliubov interferometer \add{also} enables a resonant enhancement of an arbitrarily small Andreev reflection probability. I demonstrate this with a slightly different setup, depicted in Fig.~\ref{Fig3}a. Here, the interferometer is formed by just one constriction, that encircles the bottom edge channel near the S lead. This is a slightly modified version of the setup of Fig.~\ref{Fig1} obtained by pinching-off the constriction 2 and moving the terminal $\rm N_2$ on the left-hand side  upstream the terminal $\rm N_1$. Hence, \add{the only non-trivial coefficients $T_{21}^{\alpha\beta}$ are equal to $T_{11}^{\alpha\beta}$ for the previous setup in the limit $R_2=1$}:
\begin{equation*}
	T_{21}^{eh(he)} = 1-T_{21}^{ee(hh)}=\left(\frac{T|t_{eh}|}{1+R-2\sqrt{R}|t_{ee}|\cos{\Phi}}  \right)^2,
\end{equation*}
where $R$ and $T$ are, respectively, the reflection and transmission probabilities of the constriction.

Fig.~\ref{Fig3}c shows a color-scale plot of $G_{21}$ in dependence of $R$ and the AB phase. Here, a relatively weak Andreev reflection is chosen $|t_{eh}|^2=0.1$. The dependence $G_{21}(\Phi)$ along the horizontal cross-cut is shown in Fig.~\ref{Fig3}d. A sharp resonant enhancement of the non-local conductance up to $G_{21}=2e^2/h$, the value that corresponds to $T_{21}^{eh(he)}=1$ is evident. In Fig.~\ref{Fig3}e, $G_{21}$ is plotted as a function of $|t_{eh}|$ and $T$ on resonance ($\Phi=0$). The most striking is the possibility to observe $G_{21}=2e^2/h$ for arbitrary small Andreev amplitude  $|t_{eh}|$, see the dark-red region outgoing from the origin. The dependence $G_{21}(|t_{eh}|)$ along the horizontal cross-cut of this plot at $T=0.3$ is shown in Fig.~\ref{Fig3}f, with maximum attained at $|t_{eh}|\approx0.2$. More rigorously, the optimum of $T_{21}^{eh(he)}=1$  occurs for $T^\mathrm{opt}=2|t_{eh}|/(1+|t_{eh}|)$, and Fig.~\ref{Fig3}b qualitatively explains the role of the interference in this effect. For $T=T^\mathrm{opt}$, the cross-Andreev amplitudes interfere constructively that results in $|s_{21}^{he}|=1$. At the same time, the normal scattering amplitude $(r)$ for the shortest path that didn't enter the interferometer is exactly canceled by all other paths $(-r)$, so that $s_{21}^{ee}=0$. \add{The observed resonant effect bares similarity with nearly quantized non-topological zero-bias peaks in Majorana-like nanowire devices~\cite{Moore_PRB_2018}. However, possible applications of the setup of Fig.~\ref{Fig3}a are much wider than this. In essence, Figs.~\ref{Fig3}e and~\ref{Fig3}f demonstrate complete tuning of the electron-hole superposition that makes single edge interfetometer an on-demand source of downstream Bogoliubov quasiparticles and can be implemented, e.g., in two-particle interference devices~\cite{Beenakker_2014,Ferraro_2015}.} Note, that without a superconductor a single edge interferometer has no effect, $|s_{21}^{ee(hh)}|\equiv1$ ~\cite{Gornyi_private}.

A possibility of fine tuning of the normal and Andreev scattering by the AB phase makes the proposed Bogoliubov interferometer 
a novel class of three-terminal NSN devices, that continue to gain interest in Cooper pair splitting~\cite{Hofstetter_2009,Herrmann2010,DasDas2012,Scheruebl_2022,Bordoloi_2022,Wang_2022}, in Majorana research~\cite{Menard_2020,puglia2021,Yu_2021,Wang_2022_PRB} and in charge-heat separation~\cite{Denisov_2021,Denisov_2022}. Below I schematize the main factors limiting the performance of the Bogoliubov interferometer. Dephasing by various inelastic processes is similar to other quantum Hall interferometers, implying that experiments should be performed at the lowest possible temperature with minimized bias voltages and environmental noises. In addition, the Bogoliubov interferometer relies on a coherence between the electron and hole components of the wave-function. Hence the bath temperature and the bias voltage should be kept below $\hbar/(N\tau)$, where $\tau$ is the time of flight around the interferometer and $N\approx 1/(T_1+T_2)$ is the number of turns, that is a factor of $N$ below the level spacing. 

Realistic quantum Hall experiments are conducted in a Tesla-range magnetic fields. The major concern in such high fields is a quasiparticle loss, most likely mediated by vortices in the s-wave type-II superconductor usually employed for contacting in a planar geometry in strong magnetic fields~\cite{Kurilovich_2022_Disorder,zhao2022loss}. The vortex cores serve as sub-gap reservoirs for quasiparticles, that makes the matrix $\hat{S}$ non-unitary and should drastically suppress the visibility in the same way as the suppression of the mesoscopic conductance fluctuations analyzed in Ref.~\cite{Kurilovich_2022_Disorder}. The fact that such fluctuations in a recent experiment~\cite{Zhao_2020} were at least two orders of magnitude smaller than $2e^2/h$ indicates a very strong quasiparticle loss. The effect of vortices can be potentially overcome by keeping the bias and temperature inside the mini-gap in the spectrum of the Caroli -- de Gennes -- Matricon states in the vortex core. Remaining in the planar geometry one could minimize the loss using type-II superconductors with the shortest possible coherence length (e.g. strongly disordered). This guaranties the smallest radius of the vortex core and the largest mini-gap. Alternatively, one could go out of the plane with a few nanometer thin type-I superconductor film deposited on a cleaved edge of a conventional III-V heterostructure. This may serve as a vortex-free S-terminal in strong magnetic fields directed along the film~\cite{Vaitikienas_2020} but perpendicular to the 2DES.

A so far ignored Coulomb interaction is another key player in quantum Hall interferometry, see Ref.~\cite{Carrega_2021} for a recent review. Charging effects~\cite{Halperin_2011}, which dominate the AB oscillations pattern in small-area Fabry-P\'erot devices~\cite{Zhang_2009}, are not so important in the present case, thanks to the S-terminal providing a charge sink. More intriguing is quasiparticle fractionalization, an effect associated with the interplay of two modes of collective excitations at the $\nu=2$ quantum Hall edge~\cite{Bocquillon_2013,Hashisaka_2017,Rodriguez_2020,Fujisawa_2022}. In analogy to the usual quantum Hall interferometers~\cite{Levkivskyi_2008,Ferraro_2017}, I expect a separation of the faster moving  charge-mode from the slower moving spin-mode  to impact the visibility of the AB oscillations in the Bogoliubov interferometer beyond the zero temperature/zero bias limit. This and other intriguing related effects~\cite{Choi_2015}, however, may be less pronounced in graphene, where Coulomb edge reconstruction is much weaker as compared to conventional 2DESs~\cite{Coissard_2023}. 

In summary, I proposed a Bogoliubov interferometer consisting of a superconducting terminal and a chiral edge channel encircled by one or two constrictions. The \add{charge and heat responses} of such a three-terminal \add{NSN interference }device exhibit \add{versatile tunability via} reflection probabilities of the constrictions and the Aharonov-Bohm phase. \add{A single edge version of the interferometer enables full on-demand tuning of the electron-hole superposition and can serve as a building block in more complex future interference devices.} Overall, such an interferometer may turn an attractive direction in superconductor -- quantum Hall research. 

I acknowledge valuable discussions with I.S. Burmistrov, I.V. Gornyi, V.D. Kurilovich, A.S. Melnikov, E.V. Shpagina and E.S. Tikhonov. I am especially grateful to V.D. Kurilovich for pointing out a mistake in the first version of the manuscript. The work is financially supported by the RSF project 22-12-00342.

\end{document}